*Article*

# Pressure influence on excitonic luminescence of $CsPbBr_3$ perovskite

**Lev-Ivan Bulyk [1,2,*], Taras Demkiv [2], Oleh Antonyak [2], Yaroslav M. Chornodolskyy [2], Roman Gamernyk [2], Andrzej Suchocki [1,*], Anatolii Voloshinovskii [2]**

[1] Institute of Physics, Polish Academy of Sciences, al. Lotników 32/46, 02-668, Warsaw, Poland
[2] Ivan Franko National University of Lviv, 8 Kyryla i Mefodiya St., Lviv, Ukraine
* Correspondence: suchy@ifpan.edu.pl (A.S.), bulyk@ifpan.edu.pl (L.-I.B.)

**Abstract:** This study investigates the effect of hydrostatic pressure on the luminescence properties of $CsPbBr_3$ single crystals at 12 K. The luminescence at the edge of the band gap reveals a structure attributed to free excitons, phonon replica of the free excitons, and Rashba excitons. Changes in the relative intensity of the free and Rashba excitons were observed with increasing pressure, caused by changes in the probability of nonradiative deexcitation. At pressures around 3 GPa, luminescence completely fades away. The red shift of the energy position of the maximum luminescence of free and Rashba excitons in pressure ranges of 0-1.3 GPa is attributed to the length reduction of Pb–Br bonds in $[PbBr_6]^{4-}$ octahedra, while the high-energy shift of the Rashba excitons at pressures above 1.3 GPa is due to $[PbBr_6]^{4-}$ octahedra rotation and changes in the Pb–Br–Pb angle.

## 1. Introduction

Halide perovskites $CsBX_3$ (B = Pb, Sn,...; X = Cl, Br, I) have gained significant attention and extensive research due to their diverse practical applications [1, 2, 3, 4, 5, 6, 7, 8, 9, 10, 11]. The fluorescence properties of these materials are of utmost importance for numerous applications, and understanding the nature of fluorescent transitions plays a crucial role in controlling the parameters of perovskite-based fluorescent materials. High-pressure investigations of perovskites provide additional insights into the underlying fluorescent processes and enable predictions of their unknown properties. Several studies employing hydrostatic pressure have been conducted on mono- and nanocrystals of halide perovskites, revealing isostructural phase transitions [12, 13] and amorphization [12, 14]. Various high-pressure studies have been performed on $CsPbBr_3$ in the single crystal [8, 14, 15, 16], powder [8, 12, 17 18], and other forms [16], consistently confirming the occurrence of phase transitions under high pressure. As pressure increases, $CsPbBr_3$ exhibits two isostructural orthorhombic phases: the first within the pressure range of 0 GPa to ~1 GPa, and the second from ~1 GPa to ~2 GPa. The pressures corresponding to the phase transitions exhibit slight variations among different studies. The two isostructural phases differ primarily in certain lattice parameters, such as the Br-Pb distance and the Br-Pb-Br angle, and the rate at which these parameters change with pressure.

According to multiple studies [12, 14, 15, 17, 18], the amorphization of $CsPbBr_3$ is observed to commence at pressures around 2 GPa. However, a contradictory finding is presented in a recent publication [19], where the authors claim the absence of amorphization in $CsPbBr_3$. Instead, they propose a structural phase transition to an alternative orthorhombic Pnma structure, denoted as Pnma_2. The previous observations of amorphization were explained in the study [19] attributing them to the agglomeration of grain boundaries in powder samples, including nanoparticles, nano-plates, and other types of powdered $CsPbBr_3$ samples.

At liquid helium temperature, the near-bandgap luminescence of $ScPbBr_3$ exhibits the presence of free excitons (direct transitions), their phonon replicas, and the luminescence originating from Rashba excitons (indirect transitions) [20, 21]. However, at room temperature, only two partially merged bands are observable. The emission of Rashba excitons occurs due to the appearance of an indirect band gap resulting from the interaction between electrons in the conduction band and the crystal's gradient electric fields. In the one-dimensional Rashba model, the electron's motion perpendicular to the electric field induces energy splitting, leading to the formation of a doubled-minima energy zone in a one-dimensional k-space [22]. In the case of two dimensions, this energy surface takes the shape of a trough, resembling the rotation of two axis-symmetrical parabolic curves, often referred to as a "Mexican hat shape" in literature [22]. A distinctive characteristic of transitions from such states is the circular polarization of emitted radiation, serving as evidence for the presence of the Rashba effect.

The formation of gradient electric fields in organic perovskites can be attributed to the dipole moment of the organic cation. In contrast, inorganic perovskites offer various possibilities for the generation of gradient electric fields. For instance, the presence of a bromine vacancy in the $[PbBr_6]^{4-}$ octahedron or any other point defect [23] can create a non-zero electric field gradient. Additionally, gradient electric fields can arise at domain walls [24], where the perovskite structure is disrupted, resulting in the loss of the center of symmetry. Another mechanism for the emergence of internal fields is asymmetrical lattice vibrations [25]. In a study [26], the authors investigated the polarization-dependent luminescence of $CsPbBr_3$ and $CH_3NH_3PbBr_3$, concluding that the Rashba effect in $CsPbBr_3$ exhibits a dynamic nature, originating from lattice distortion occurring on a timescale of 100 fs. While one study [26] suggests the absence of Rashba splitting in $CsPbBr_3$ at low temperatures, another study [27] provides solid evidence for the presence of the Rashba effect under such conditions. In [27], the authors propose that the Rashba effect stems from lattice distortion induced by $Cs^+$ motion or surface effects in nanoscale nanocrystals.

The Rashba line in the spectra can have alternative interpretations, such as excitons bound to defects. In a study of $FA_xMA_{1-x}PbI_3$ [28], multiple lines observed below the free exciton were assigned to various types of defects. However, the objective of the current work is not to determine the exact origin of a specific line in $CsPbBr_3$, whether it is attributed to the Rashba effect or localized excitons. Rather, the focus is on investigating the luminescence dependence at low temperatures as a function of pressure.

To the best of our knowledge, previous high-pressure measurements of halide perovskites have predominantly been carried out at room temperature. However, it is crucial to note that the luminescence characteristics of $CsPbBr_3$ display a more pronounced structure at lower temperatures when compared to room temperature. Therefore, conducting investigations on the impact of pressure on luminescence at lower temperatures holds significant importance as it can provide valuable insights into the fundamental nature of near-bandgap luminescence in halide perovskites.

To investigate the excitonic lines of $CsPbBr_3$ single crystals, luminescence studies were conducted within the pressure range of 0–4.2 GPa. To distinguish free excitons from their phonon replicas and Rashba excitons, these measurements were carried out at liquid helium temperature. The use of such low temperatures allows for a clearer resolution of the different luminescence components.

## 2. Materials and Methods

Single crystals of $CsPbBr_3$ were synthesized using the vertical Bridgman method. CsCl and $PbBr_2$ were preliminarily cleaned using the zone melting procedure before the growth process. The structural phase of the sample was determined using XRD analysis, which showed the dominance of the orthorhombic phase, in agreement with previous findings.

High-pressure measurements were conducted using a miniature diamond anvil cell (DAC) from easyLab, which can operate at low temperatures. The sample was chopped into small plate-shaped pieces with a thickness of approximately 25-28 μm and a size of about 200 μm in the remaining two dimensions. Ruby was used as a pressure sensor, and argon was used as the pressure-transmitting medium.

The measurements were performed in backscattering geometry using a semiconducting laser with a 405 nm emission wavelength. For low-temperature measurements, the sample was loaded into the continuous flow CF200 cryostat provided by Oxford Instruments. Spectra were resolved using a Triax 320 monochromator produced by ISA Yobin Yvon-Spex, equipped with a Spectrum One liquid nitrogen-cooled CCD camera. The spectral dispersion of the system was better than 0.3 nm.

## 3. Results and discussion

The luminescence spectrum of $CsPbBr_3$, measured at 12 K and ambient pressure, is depicted in Fig. 1. The line corresponding to free excitons is observed at 534.0 nm, while the phonon replica of free excitons ($h\omega_{LO}$ = 19 meV [29, 30]) appears at 536.5 nm. Additionally, the line at 539.7 nm can be assigned to defect-bound excitons. However, the interpretation of the line at 542 nm, as mentioned in the introduction, is still subject to debate and lacks clarity in the literature. This study primarily focuses on investigating the pressure dependence of luminescence at low temperatures, and the results obtained do not provide a definitive answer regarding the origin of the 542 nm line. Consequently, the detailed discussion concerning the interpretation of the 542 nm line is presented in the Supplementary Information, where we provide some preliminary evidence supporting the interpretation of the Rashba effect. For the sake of clarity in the discussion, the assignment of the 542 nm line to the Rashba effect is adopted, as it is not pivotal for further considerations.

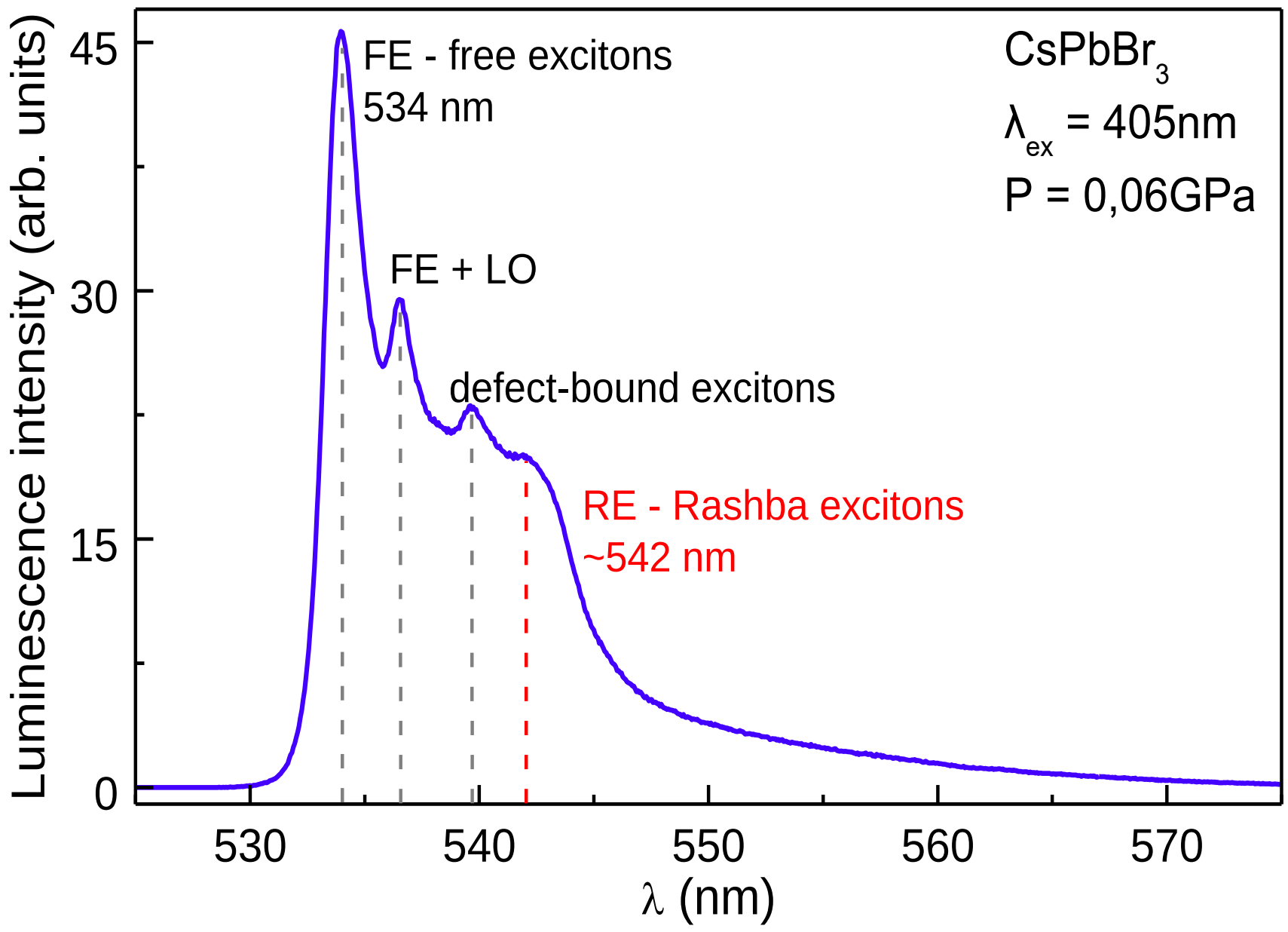

**Fig. 1.** Luminescence of $CsPbBr_3$ at the lowest pressure (0,06 GPa) and low temperature (T = 12 K).

The luminescent spectra as a function of pressure are presented in Fig. 2. The spectra are divided into two partially covered pressure ranges: (a) 0.06—1.53 GPa and (b) 0.98—4.22 GPa. The decision to present the spectra in two separate graphs aims to enhance readability and facilitate the visualization of the transition from the first pressure region to the second pressure region. By partially covering the spectra, it becomes more apparent how the luminescence characteristics evolve between the two pressure regions.

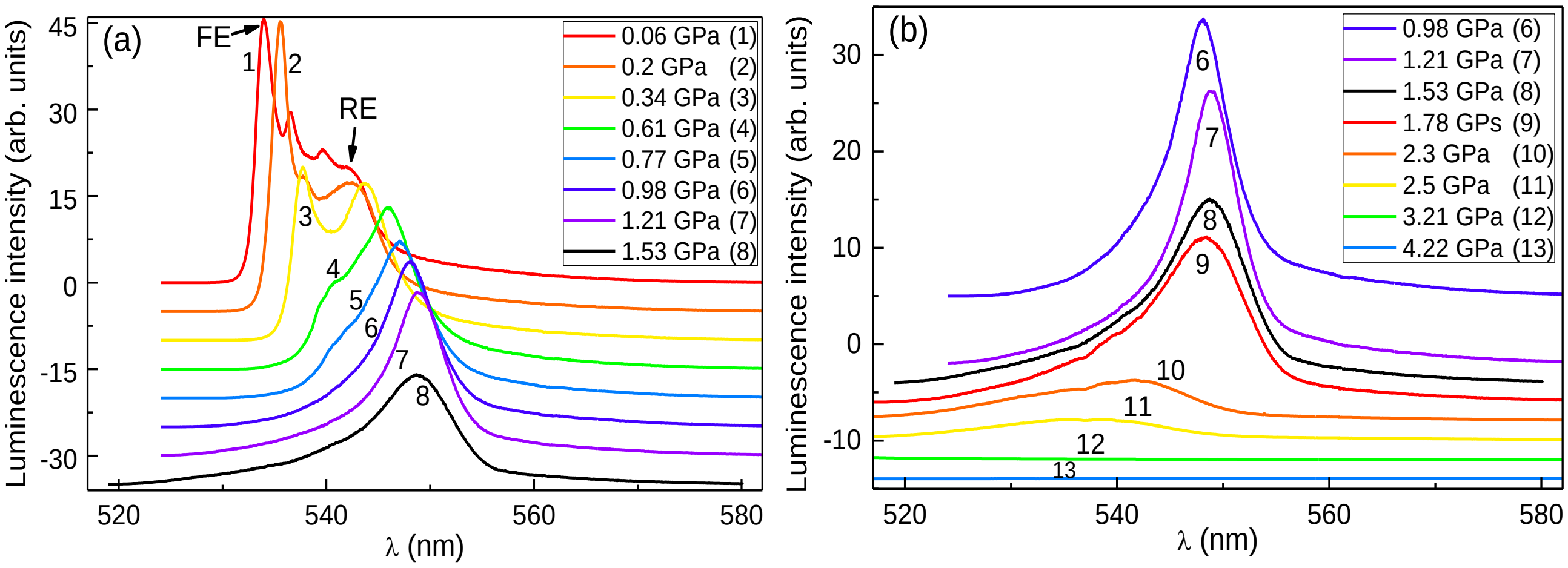

**Fig. 2.** Dependence of the $CsPbBr_3$ luminescence on pressure at T = 12 K. For better readability, the luminescence spectra are presented in two separate, partially overlapped pressure regions: **(a)** – 0.06 – 1.53 GPa and **(b)** – 0.98 – 4.42 GPa.

The model of edge luminescence, which incorporates both direct and indirect exciton bands, explains the luminescence spectra behavior of $CsPbBr_3$ under high pressure. The luminescence spectra, as a function of pressure, exhibit two distinct groups of peaks associated with direct band excitons (534 nm) and indirect band (Rashba) excitons (542 nm). The intensity of the luminescence from free excitons and their phonon replica (Fig. 2a, peak 534.0 nm — free excitons (FE)) decreases as pressure increases. At a pressure of 1 GPa, the luminescence from free excitons and their phonon replica appears to be nearly extinguished (as seen in Fig. 2), forming a featureless shoulder adjacent to the Rashba line. In contrast, the intensity of the luminescence from Rashba excitons (Fig. 2a, peak 542 nm, RE) increases with pressure. However, at pressures exceeding 1.3 GPa, the intensity of the Rashba excitons starts to decrease significantly and eventually disappears completely at a pressure of 4.2 GPa.

The integral intensity of the $CsPbBr_3$ luminescence was determined by calculating the area under the luminescence curves. The relationship between the integral intensity and pressure is depicted in Fig. 3a, represented by black squares and a black eye-guiding line. The intensity of the Rashba emission was obtained from the peaks' maxima observed in Fig. 2 for all spectra, and it is also displayed in Fig. 3a as red circles along with a red eye-guiding line. The intensity of the free excitons was estimated by subtracting the Rashba intensity from the integral intensity, and these values are presented in Fig. 3a as blue triangles along with a blue eye-guiding line. It should be noted that this estimation of the free excitons' intensity yields slightly higher values since it includes the intensity of the phonon replicas of the free excitons.

The presence of defect luminescence in the spectra may as well introduce some imperfections in the approach used. However, we believe that the impact of defect luminescence is relatively small since the defect bands are primarily located in the longer wavelength range (550-620 nm), as shown in Fig.SI. 1 in the supplementary information. While some excitons localized on defects may be present in proximity to the Rashba line (around 540 nm, as observed in Fig.SI. 2), the subsequent discussion on energy redistribution processes between free and Rashba excitons can also be extended to include defect-bound excitons. Furthermore, the Supplementary information provides a comparison between the results obtained using this approach and the results from the direct fitting of the spectra with Gaussian functions.

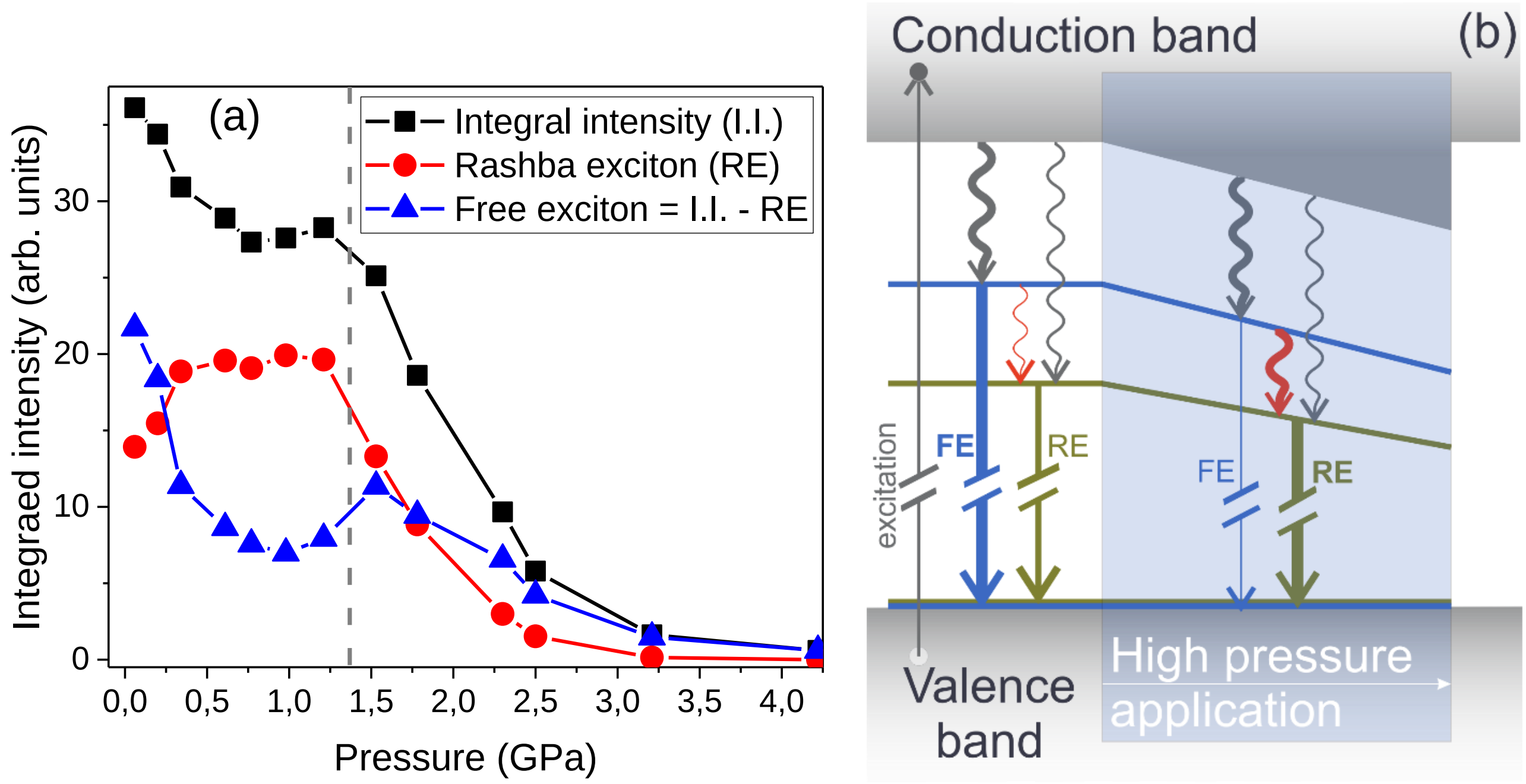

**Fig. 3. (a)** Pressure dependence of the integral intensity of $CsPbBr_3$ luminescence, and pressure dependence of the in-peak intensity of the Rashba excitons emission. **(b)** Schematic representation of the position of free and Rashba excitons on pressure.

From Fig. 2, it can be observed that the intensity of the free exciton at 0.2 GPa is slightly higher than that at 0.06 GPa. In a previous work [19], the authors demonstrated an increase in luminescence intensity with increasing pressure up to approximately 90 times at room temperature. They attributed this effect to the reduction in the distance between the free and Rashba excitons under pressure. In our case, as shown in Fig.SI. 1 in the supplementary information, the luminescence intensity at low temperatures is approximately four orders of magnitude higher compared to room temperature. Therefore, we may be observing a similar effect as reported in [19], albeit weaker, possibly due to the higher photoluminescence quantum yield at low temperatures or the lower quality of our sample.

The increase in integral intensity within the pressure range of 0.75 GPa to 1.25 GPa is not clearly understood. However, it can potentially be explained using similar reasoning as mentioned above, following the explanation provided in [19].

As depicted in Fig. 3a, the luminescence of free excitons and their phonon replicas is quenched with increasing pressure up to 1 GPa. Conversely, the luminescence of Rashba excitons gains intensity within this pressure range. At higher pressures, both the free and Rashba excitons are quenched. It appears that free excitons and Rashba excitons compete as deexcitation channels for excited electron-hole pairs created during the excitation processes (see Fig. 3b).

The energy difference between free excitons and Rashba excitons is about 35 meV (~350 $cm^{-1}$) and decreases with the applied pressure to about 25 meV (~202 $cm^{-1}$) at a pressure of about 1 GPa (see Fig. 4). The decrease in the intensity of free excitons and the increase in the intensity of Rashba excitons with the increase in pressure can be explained by the increased probability of nonradiative deexcitation from free excitons to Rashba excitons (schematically shown in Fig. 3b). This process can be explained by the "energy-gap law" often observed in rare-earth spectroscopy [31, 32]. According to this law, the multiphonon emission rate $A_{nr}$ between two states separated by the energy $\Delta E$ is proportional to the exponent of the energy $\Delta E$:

$$A_{nr} = A_{nr}(0)e^{-\alpha \Delta E} \quad (1)$$

where $A_{nr}(0)$, and $\alpha$ are characteristic parameters for the material. The energy gap $\Delta E$ may be expressed as:

$$\Delta E = n \cdot \hbar\Omega \quad (2)$$

where $\hbar\Omega$ is the effective phonon energy, and $n$ is the number of phonons needed to cover the energy distance $\Delta E$. Considering the energy distances between free excitons and Rashba excitons at ambient pressure and about 1 GPa and the LO phonon energy of $CsPbBr_3$ (19 meV = 155 $cm^{-1}$ [29, 30]), the ratio of nonradiative transition rate between these levels at ambient pressure ($A_{nr}(amb.pressure) = A_{nr}(0)e^{-\alpha(2\cdot\hbar\Omega)}$) and 1 GPa ($A_{nr}(1\,GPa) = A_{nr}(0)e^{-\alpha(1\cdot\hbar\Omega)}$) is approximately $e^1 \approx 2.72$ times. This indicates that the probability of nonradiative deexcitation between the free exciton and Rashba exciton increases by approximately 2.7 times between ambient pressure and 1 GPa. We make these estimations based on the assumption that the effective phonon energy and the electron-phonon coupling remain constant under pressure. It should be noted that the effective phonon energy may be lower than 155 $cm^{-1}$, considering the characteristic low-energy TO phonons of $CsPbBr_3$ (with an energy of about 75 $cm^{-1}$) [32, 33]. This could further contribute to the increase in the nonradiative transition probability between free and Rashba excitons due to the pressure-induced change in the energy difference between these quasi-particles.

The increase in the intensity of the luminescence band at 542 nm (Fig. 3a) with increasing hydrostatic pressure can be used as an additional marker of the assignment of this band to Rashba excitons.

Lastly, the probability of nonradiative deexcitation from free excitons to Rashba excitons decreases above 1 GPa, leading to the increase in the intensity of free excitons (see Fig. 3a). The reappearance of free excitons (although much broadened) reflects in the emission spectra as a small high-energy shoulder at pressures higher than 1.3 GPa (see Fig. 2b).

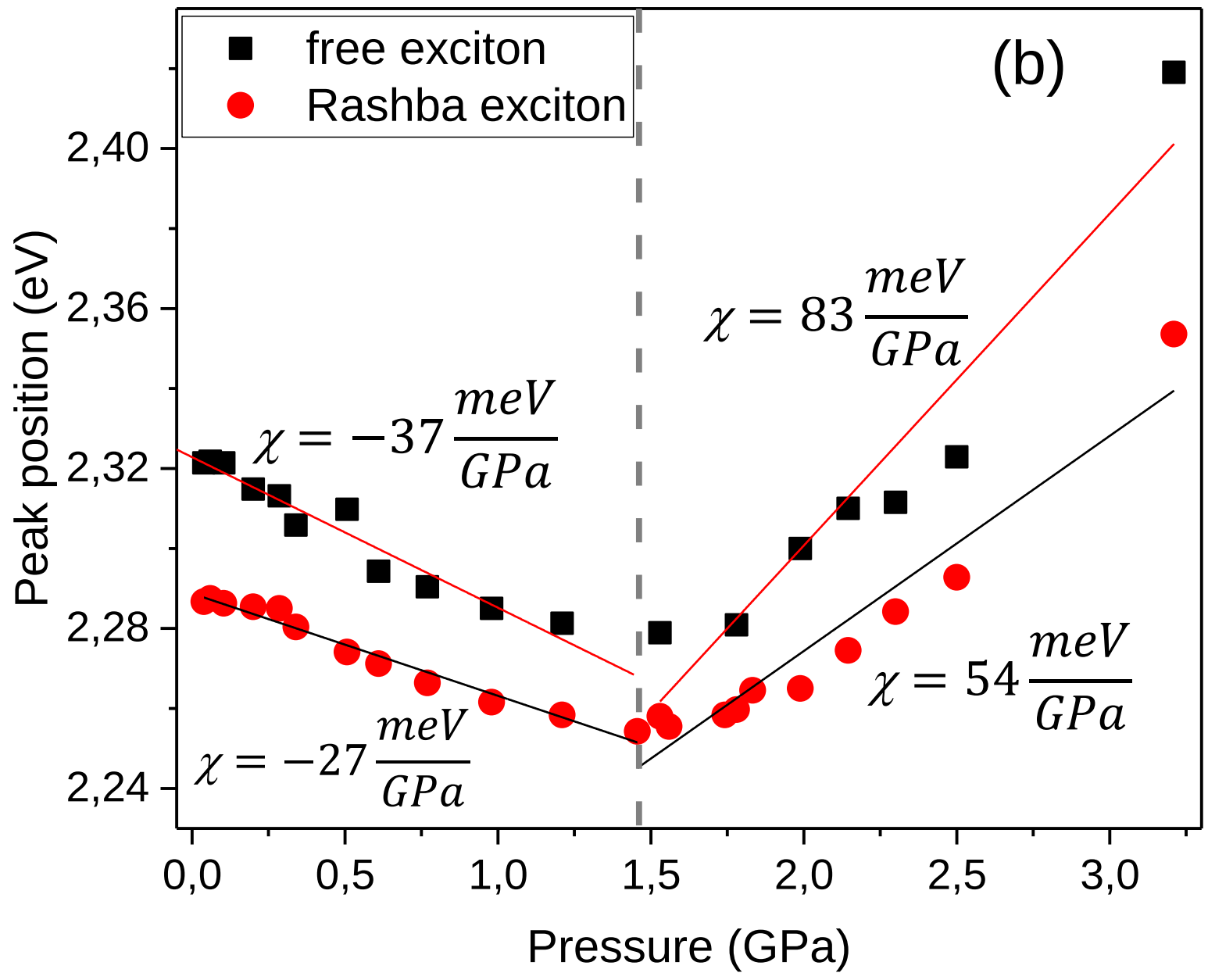

**Fig. 4.** Dependence of the position of $CsPbBr_3$ luminescence maxima on pressure at T = 12 K.

Fig. 4 shows the dependence of the energy position of the luminescence maxima of direct and indirect excitons on pressure. In the pressure range of 0–1.3 GPa, a shift of the energy positions of the luminescence maxima towards the region of low energies (in the long-wavelength direction) is observed. After 1.3 GPa, a reverse shift towards higher energies occurs. The observed shifts in luminescence can be explained by the changes of the band gap structure of $CsPbBr_3$, which are caused by the phase transitions. The observed shifts in luminescence can be mathematically described by the dependence adopted from [34]:

$$E_g = E_{opt} + \chi p \quad (3)$$

where $E_{opt}$ is the ambient optical gap, $\chi$ is the so-called pressure coefficient that indicates the change in the band gap (in meV) with a hydrostatic pressure change of 1 GPa. The fitting by equation (3) is shown in Fig. 4. The obtained values of coefficients are $\chi$ = -37 meV/GPa and $\chi$ = -27 meV/GPa for the free and Rashba excitons, respectively, in the pressure range of 0–1.3 GPa, and $\chi$ = 54 meV/GPa and $\chi$ = 83 meV/GPa for the free and Rashba excitons, respectively, for pressures above 1.3 GPa. The values coincide with the results of previous studies for single crystals and nanoparticles of $CsPbBr_3$ at room temperature [12, 13, 14, 17, 18]. The dependence of the luminescence maxima on pressure is consistent with the structural (phase transitions are known to occur under high pressure in $CsPbBr_3$ [8, 12, 13, 14, 15]) and band gap changes that occur in the crystals when pressure is applied. Several ranges of pressure with characteristic changes in luminescence parameters can be distinguished.

At pressures of 0–1.3 GPa, when a shift of the luminescence bands towards the region of low energies is observed, there is a decrease in the parameters of the crystal lattice, a decrease in the lead–halide cation distance, and a decrease in the width of the band gap. To qualitatively explain the change in the band gap width due to pressure, the band-gap model using the molecular orbital method can be used. In [35], the authors state that the bottom of the conduction band is determined by the nonbonding localized state of the *6p* orbital since the density of *6p* states of lead is predominant. In the case of the non-binding nature of the bottom of the conduction band, the effect of pressure is not significant for the energy position of the bottom of the conduction band [35]. Consequently, the width of the band gap is determined by the features of the energy states of the valence band.

The top of the valence band exhibits an antibonding character due to the interaction between the *6s* Pb and *4p* Br atomic orbitals, primarily influenced by the *4p* Br orbitals [35, 36]. The application of pressure enhances the overlap of wave functions within the valence band, causing a weakening of the bond due to its antibonding nature [12]. Consequently, the top of the valence band shifts towards higher energy regions, leading to a reduction in the band gap.

In their work [12], the authors present theoretical calculations for pressures higher than 1.3 GPa. They conclude that the inclinations of the octahedra and the resulting changes in the angles of the Pb-Br-Pb bonds are decisive for the change in the band gap. Calculations show that in the pressure range from 1.3 GPa to 4.2 GPa, the Pb-Br-Pb angle changes from 153° to 133° [35], and the Pb-Br distance increases. Such changes lead to a weakening of the connection between the *6s* Pb and *4p* Br electron clouds, which, given the antibonding nature of the top of the valence band, leads to a decrease in the energy of the antibonding *6s-4p* orbital and an increase in the band gap. These changes in the band gap are consistent with the behavior of the position of the luminescence maxima as a function of pressure.

From Fig. 4, it can be seen that the energetic distance between the free and Rashba excitons is not constant with pressure. It decreases and increases respectively below and above 1.3 GPa. This suggests that the Rashba splitting is decreasing and afterward increasing respectively below and above 1.3 GPa. In their work [21], the authors report that the Rashba splitting decreases with the increase in pressure for $MAPbI_3$, which supports our observation. As discussed above, below 1.3 GPa, the intensity of the Rashba excitons is increasing due to the increase in the probability of nonradiative deexcitation from the free excitons to the Rashba excitons. Therefore, a small increase in the intensity of free excitons in the pressure range of 1.0–1.5 GPa can be explained by the decrease in the probability of nonradiative deexcitation caused by the increase in the Rashba splitting above 1.3 GPa.

The Rashba excitons have the lowest relative intensity compared to free excitons at 1 GPa, while the smallest Rashba splitting is at 1.3 GPa, which are slightly different pressures. We believe that this discrepancy originates from the low accuracy of the fittings shown in Fig. 4. As seen in the figure, the position dependencies of the luminescence peaks on pressure are not linear, but rather follow parabolic curves. Straight-line fittings were used because fitting with parabolas does not have a physical background. The parabolic-like dependence of the $CsPbBr_3$ luminescence position on pressure is reported in many other works, both experimental and theoretical [12, 14, 34, 35].

Another perspective worth discussing thing is that the luminescence behavior under pressure can be attributed to structural phase transitions occurring at high pressures. These phase transitions can lead to changes in the direction of the shift observed in the emission lines. However, several experimental studies [8, 12, 14, 15, 16, 17, 18] have demonstrated that the phase transitions are isostructural, indicating that the principal phase remains unchanged. Therefore, it is concluded that isostructural phase transitions alone are not sufficient to explain the luminescence behavior observed under pressure.

## 4. Conclusions

The edge luminescence spectra of $CsPbBr_3$ single crystals at 12 K revealed distinct features, including a narrow luminescence line corresponding to free excitons (direct transitions, 534.0 nm), a phonon replica line (534.5 nm), and a narrow band (542 nm) attributed to Rashba excitons (indirect transitions). However, the nature of the 542 nm band is still unclear, and an alternative possibility is that it arises from excitons localized on a defect. The energy of the phonon replica for free excitons coincided with the longitudinal optical phonon energy ($h\omega_{LO}$ = 19 meV [29, 30]).

This study aimed to investigate the effect of high pressure on the edge luminescence of $CsPbBr_3$ single crystals at a temperature of 12 K. The results revealed distinct behavior in the luminescence intensity of direct and indirect excitons under pressure, providing additional evidence for their different nature. The changes observed in the intensity of Rashba excitons can be attributed to variations in the probability of nonradiative deexcitation from free excitons to Rashba exciton states.

As the pressure increases up to 1.3 GPa, the Rashba splitting decreases, resulting in an increased intensity of Rashba excitons due to enhanced nonradiative deexcitation from free to Rashba exciton. Conversely, as the pressure continues increasing above 1.3 GPa, the Rashba splitting increases, leading to an increased intensity of free excitons. The discrepancy in pressure values where the Rashba splitting is minimal (1.3 GPa) and where the intensity of Rashba excitons is lowest (1 GPa) can be attributed to the limited precision of fitting the non-linear dependencies with straight lines.

Within the pressure range of 0-1.3 GPa, the emission maxima of direct and indirect excitons shifted towards the low-energy region. Conversely, at pressures above 1.3 GPa, the bands of indirect excitons shifted towards the high-energy region. These observations align with changes in the band gap energy resulting from structural modifications, including Pb-Br bond shortening and octahedra inclination through the variation of the Pb-Br-Pb bond angle.

Finally, at the pressure about 3 GPa, the luminescence of $CsPbBr_3$ was completely quenched. This phenomenon could be attributed to either amorphization processes occurring in the sample at higher pressures or another phase transition to a non-emissive phase.

**Supplementary Materials:** The Supporting Information can be downloaded at: {reference here}.

**Author Contributions:** Conceptualization, T. Demkiv, A. Suchocki, and A. Voloshinovskii; validation, A. Suchocki and A. Voloshinovskii; investigation, L.-I. Bulyk; data curation, L.-I. Bulyk; writing—original draft preparation, T. Demkiv, O. Antonyak, Ja. Chornodolskyy, and R. Gamernyk; sample preparation, O. Antonyak, R. Gamernyk; writing—review and editing, A. Suchocki, L.-I. Bulyk; supervision, A. Suchocki, and A. Voloshinovskii. All authors have read and agreed to the published version of the manuscript.

**Declaration of generative AI and AI-assisted technologies in the writing process**: During the preparation of this work the author(s) used ChatGPT in order to improve English. After using this tool/service, the author(s) reviewed and edited the content as needed and take(s) full responsibility for the content of the publication.

**Acknowledgments:** This work was partially supported by the National Science Centre, Poland, grant SHENG 2 number: 2021/40/Q/ST5/00336.

**Conflicts of Interest:** The authors declare no conflict of interest

# Supplementary information to "Pressure influence on excitonic luminescence of CsPbBr3 perovskite"

**Lev-Ivan Bulyk [1,2,*], Taras Demkiv [2], Oleh Antonyak [2], Yaroslav M. Chornodolskyy [2], Roman Gamernyk [2], Andrzej Suchocki [1,*], Anatolii Voloshinovskii [2]**

[1] Ivan Franko National University of Lviv, 8 Kyryla i Mefodiya St., Lviv, Ukraine
[2] Institute of Physics, Polish Academy of Sciences, al. Lotników 32/46, 02-668, Warsaw, Poland
[*] Correspondence: suchy@ifpan.edu.pl (A.S.), bulyk@ifpan.edu.pl (L.-I.B.)

## 1. Luminescence as a function of temperature and different points of the sample

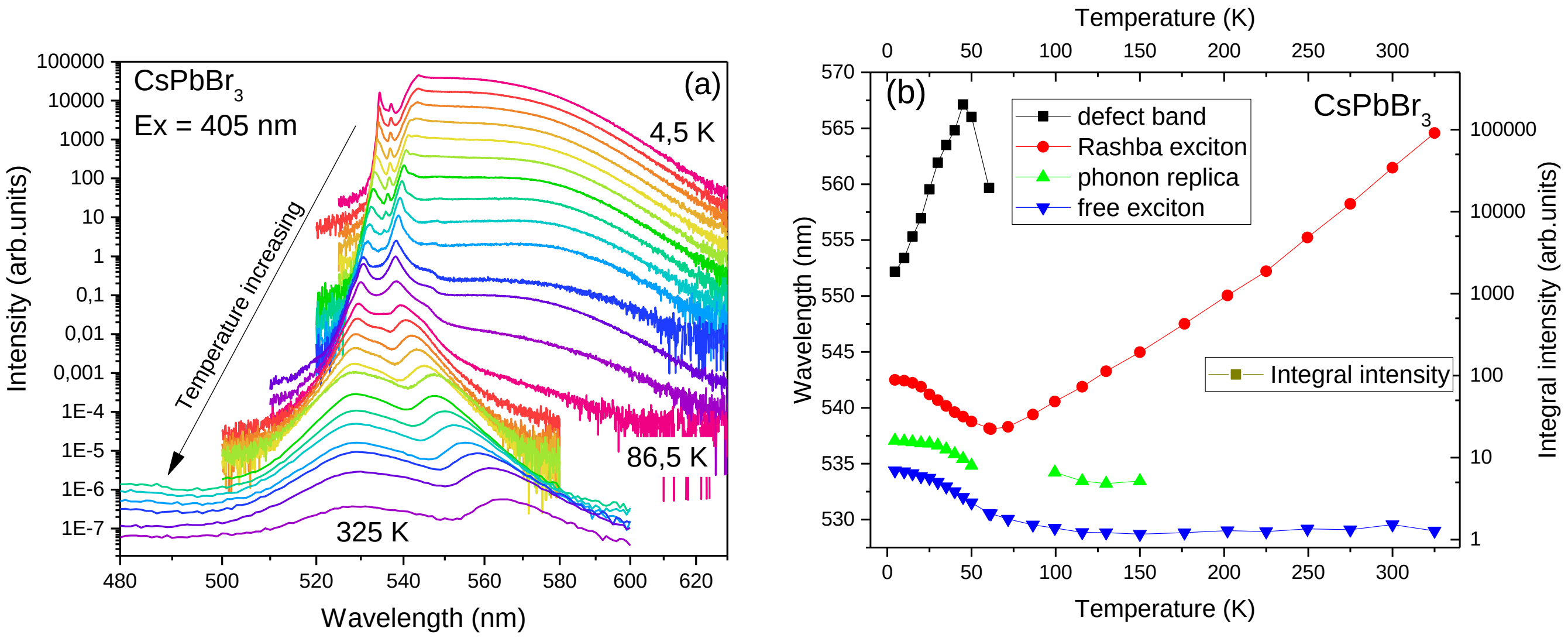

**Fig.SI. 1.** $CsPbBr_3$ luminescence dependence on temperature. (a) – luminescence spectra, (b) – position of the lines and the integral intensity of the luminescence.

In most cases, excitons bound to defects exhibit a much faster temperature-induced quenching compared to free excitons. For instance, in the study of $FA_xMA_{1-x}PbI3$ [37], various spectral lines below the free exciton were attributed to different defect types, and the luminescence of these defects was observed to decay well below 100 K. In our investigation, as depicted in Fig.SI. 1, the line assigned to the Rashba exciton demonstrates a quenching rate similar to or even slower than that of the free exciton. Hence, despite the absence of substantial evidence supporting the Rashba effect assignment in this study, it was designated as the Rashba exciton. It is worth noting that the temperature dependence of the Rashba exciton's position aligns with the findings of [38]. In contrast, the authors of [38] concluded that $CsPbBr_3$ lacks Rashba splitting at temperatures below 90 K, contradicting our observations. However, the work [39] provides compelling evidence supporting the presence of the Rashba effect in $CsPbBr_3$ at low temperatures. In conclusion, the assignment of the 542 nm line remains contentious, necessitating further investigations to address this issue.

The low-temperature spectra of the sample's luminescence are presented in Fig.SI. 2. These spectra were obtained by exciting the sample with a laser focused on various points. All spectra are normalized to the intensity of the free exciton. As depicted in Fig.SI. 2, different regions of the sample exhibit slightly different luminescence spectra, indicating sample inhomogeneity. The presence of the Rashba effect relies on the breaking of inversion symmetry within the $CsPbBr_3$ crystal structure. In the orthorhombic $CsPbBr_3$, lead (Pb) occupies a site that possesses inversion symmetry. However, if the sample is nonhomogeneous or contains defects, the symmetry may be disrupted in certain regions, thereby exhibiting Rashba luminescence.

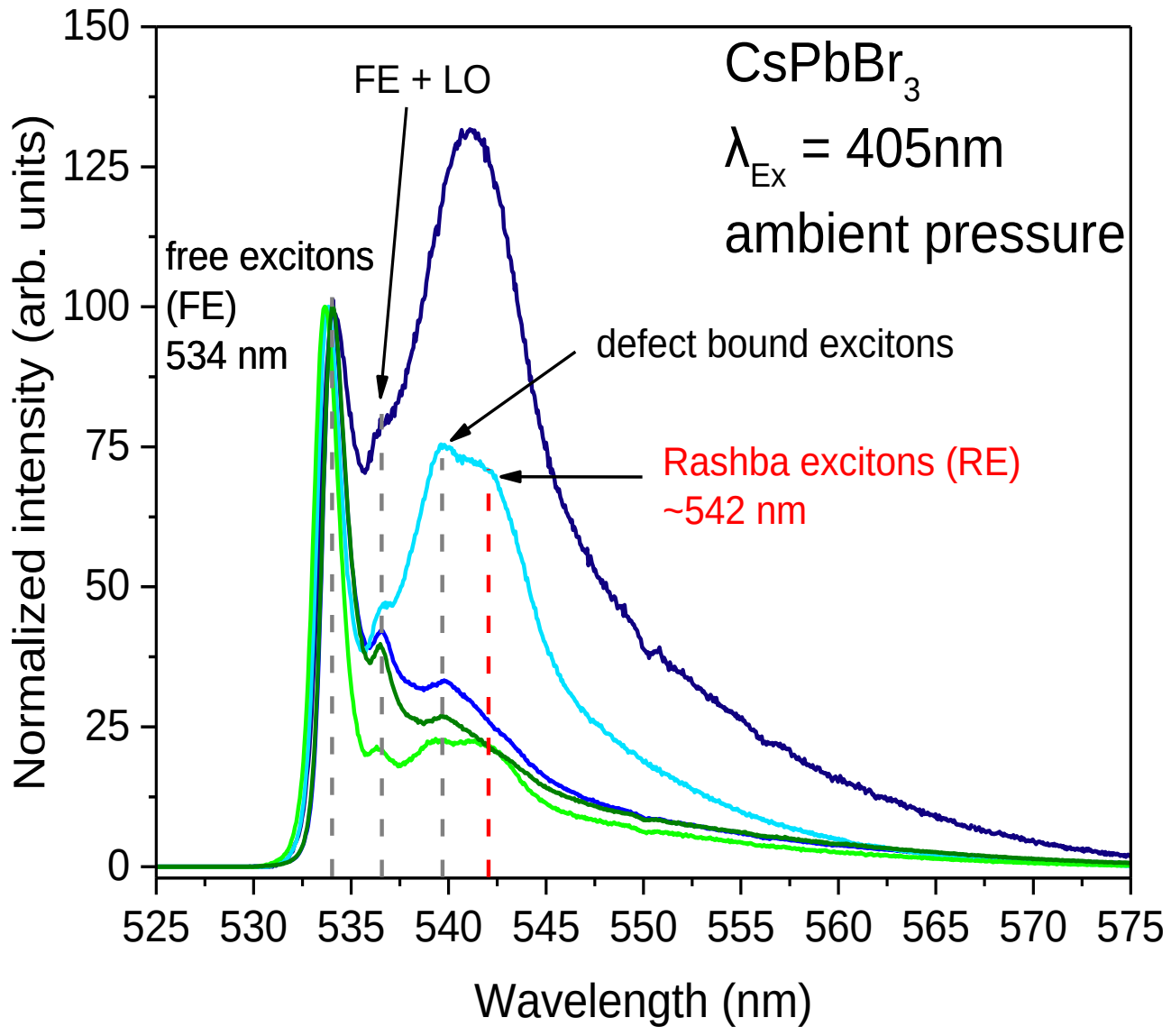

**Fig.SI. 2.** Comparison of the luminescence of different parts of the $CsPbBr_3$ sample at liquid helium temperature.

## 2. XRD Measurements of the $CsPbBr_3$

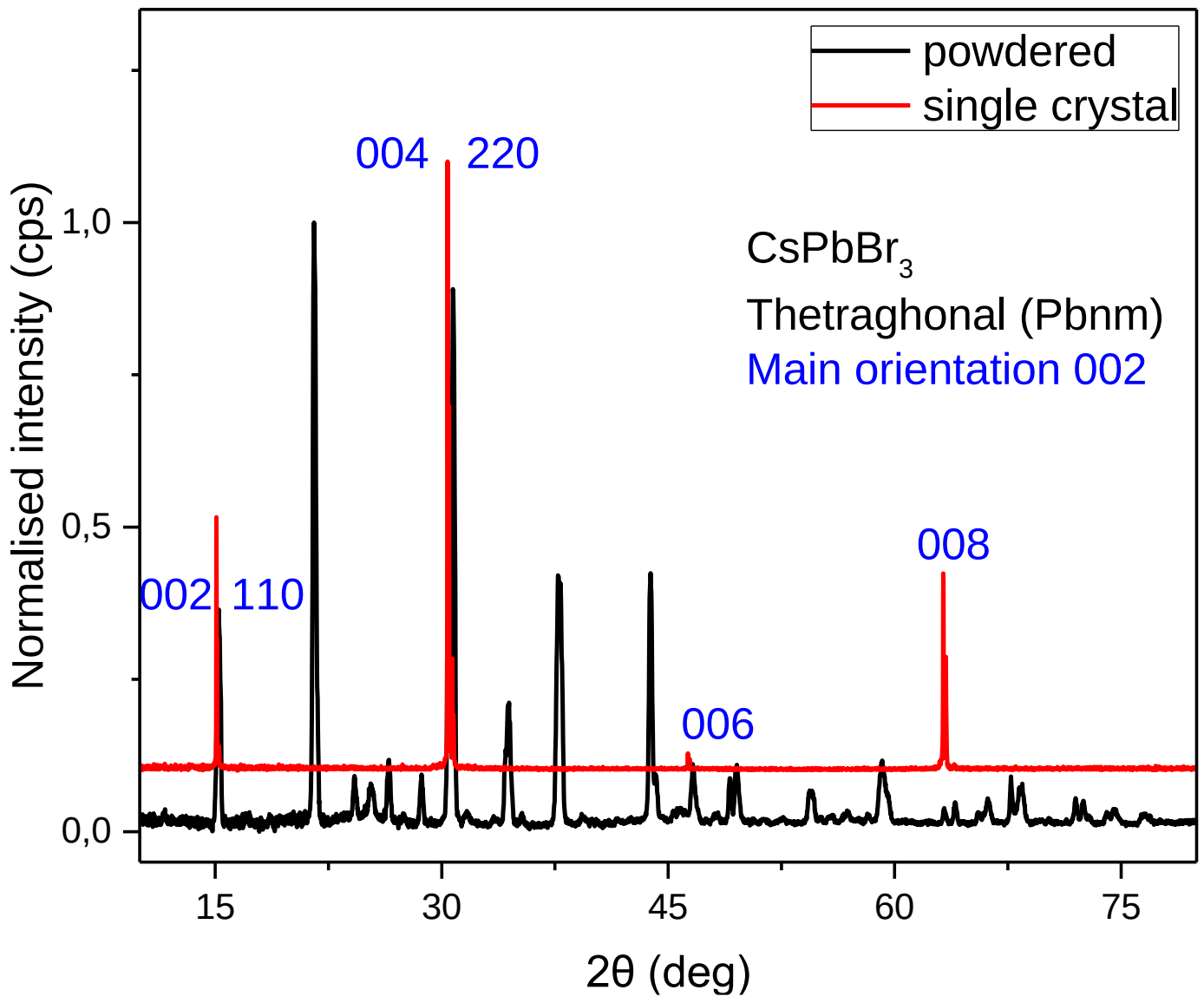

**Fig.SI. 3.** The results of the XRD measurements of single crystal and powdered $CsPBBr_3$.

As shown in Fig.SI. 3, the single crystal X-ray diffraction (XRD) measurements confirmed the presence of the orthorhombic (Pbnm) phase in the sample, with the [002] orientation. It is important to note that the crystal was not intentionally oriented prior to the measurement, resulting in a random orientation. The powder diffraction analysis also exhibited distinct diffraction lines corresponding to the orthorhombic (Pbnm) phase.

## 3. Luminescence spectra

To assess the reproducibility of the experiment, multiple measurements of $CsPbBr_3$ under pressure were conducted. The data used in this study primarily came from two sets of measurements, as shown in Fig.SI. 4. Specifically, selected curves from Fig.SI. 4a are presented in the main text. It is worth noting that the spectra in panels a and b exhibit slight differences. This variation can be attributed to the fact that the measurements were performed on two different pieces of the same inhomogeneous sample. The presence of inhomogeneity in the sample is also evident in Fig.SI. 2.

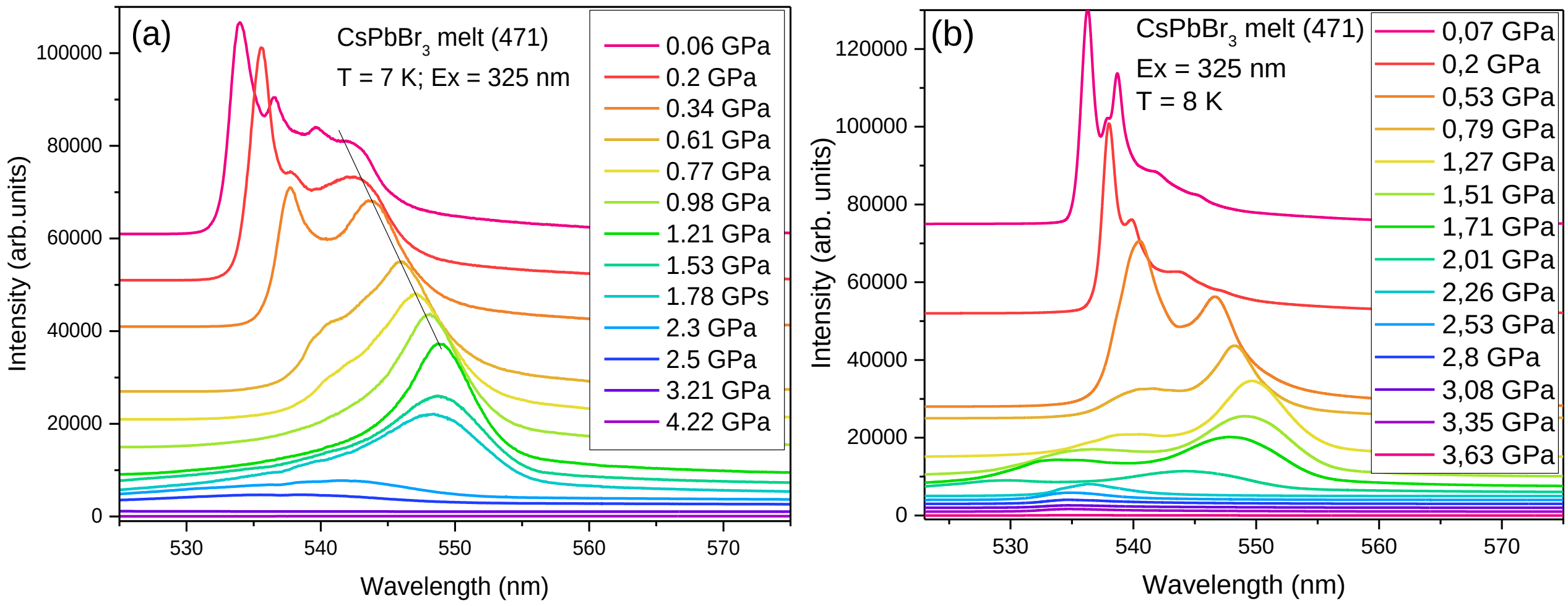

**Fig.SI. 4.** Low-temperature luminescence of $CsPbBr_3$ as a function of high pressure. (a) and (b) are two measurements of different pieces of the same sample (from the melt, number 471). Oil was used as a pressure-transmitting medium

# 4. Fittings of the spectra with Gaussian functions

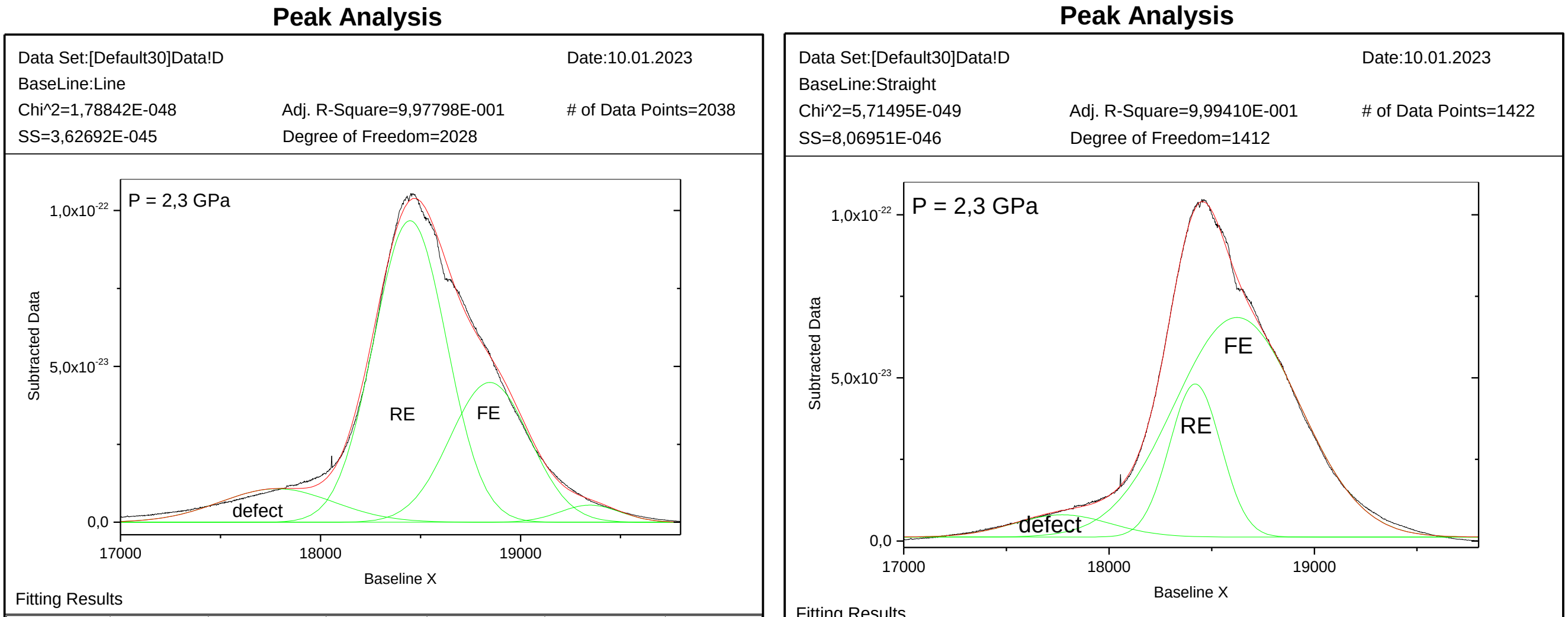

| Peak Index | Peak Type | Area Intg | FWHM | Max Height | Center Grvty | Area IntgP |
|---|---|---|---|---|---|---|
| 1 | Gaussian | 2,22939E-20 | 467,0175 | 4,48458E-23 | 18846,40624 | 29,36244 |
| 2 | Gaussian | 4,39271E-20 | 426,68473 | 9,67149E-23 | 18447,65104 | 57,85462 |
| 3 | Gaussian | 7,63371E-21 | 667,81201 | 1,07386E-23 | 17781,44455 | 10,05404 |
| 4 | Gaussian | 2,07196E-21 | 354,00394 | 5,49847E-24 | 19347,55701 | 2,7289 |

| Peak Index | Peak Type | Area Intg | FWHM | Max Height | Center Grvty | Area IntgP |
|---|---|---|---|---|---|---|
| 1 | Gaussian | 5,13328E-20 | 716,29894 | 6,73266E-23 | 18623,3177 | 73,33013 |
| 2 | Gaussian | 1,46406E-20 | 293,11812 | 4,69227E-23 | 18419,49906 | 20,91439 |
| 3 | Gaussian | 4,02897E-21 | 555,66994 | 6,81353E-24 | 17771,29117 | 5,75548 |

**Fig.SI. 5.** Two possible options for the fitting of the luminescence spectra at 2.3 GPa.

Fig.SI. 6 illustrates several luminescence spectra along with their corresponding fittings. It can be observed that the phonon replicas and defect-bound excitons diminish rapidly with increasing pressure, whereas the intensity of the Rashba exciton exhibits an opposite trend by increasing. As the pressure rises, the accuracy of the fittings decreases, leading to multiple options of slightly different Gaussian fittings. This is demonstrated in Fig.SI. 5, which showcases two different fitting options at a pressure of 2.3 GPa. Initially, they may appear significantly distinct, but when the respective parameters are inserted into the dependencies shown in Fig.SI. 7, both options yield acceptable results due to the substantial scattering of the data points.

## Peak Analysis

Data Set:[Default17]Data!C Date:10.01.2023
BaseLine:Line
Chi^2=8,18267E-047 Adj. R-Square=9,98482E-001 # of Data Points=1026
SS=8,29723E-044 Degree of Freedom=1014

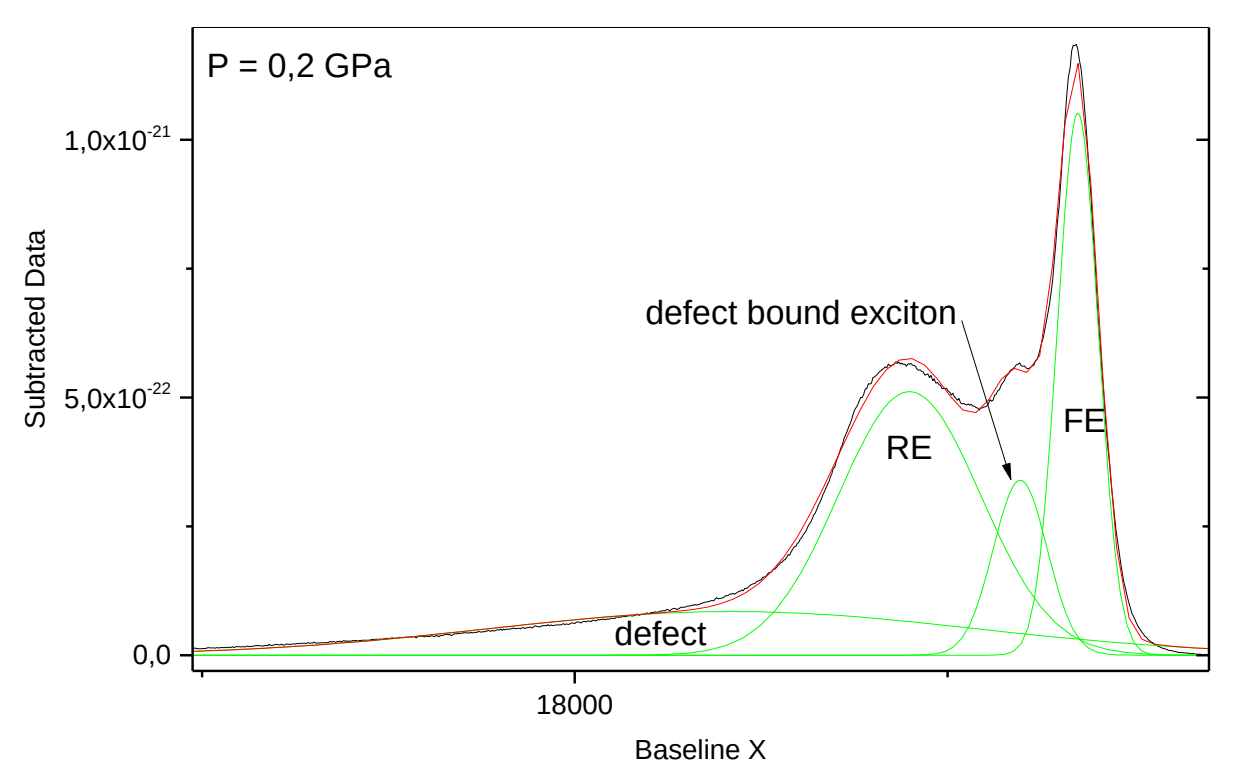

Fitting Results

| Peak Index | Peak Type | Area Intg | FWHM | Max Height | Center Grvty | Area IntgP |
|---|---|---|---|---|---|---|
| 1 | Gaussian | 7,22217E-20 | 64,53511 | 1,05133E-21 | 18674,59602 | 24,48405 |
| 2 | Gaussian | 3,08447E-20 | 85,38945 | 3,39347E-22 | 18597,42014 | 10,45674 |
| 3 | Gaussian | 1,22475E-19 | 225,14476 | 5,11036E-22 | 18448,8537 | 41,52039 |
| 4 | Gaussian | 6,94335E-20 | 770,27109 | 8,46824E-23 | 18210,71733 | 23,53882 |

## Peak Analysis

Data Set:[Default18]Data!C Date:10.01.2023
BaseLine:Line
Chi^2=2,12216E-046 Adj. R-Square=9,94907E-001 # of Data Points=1026
SS=2,15823E-043 Degree of Freedom=1017

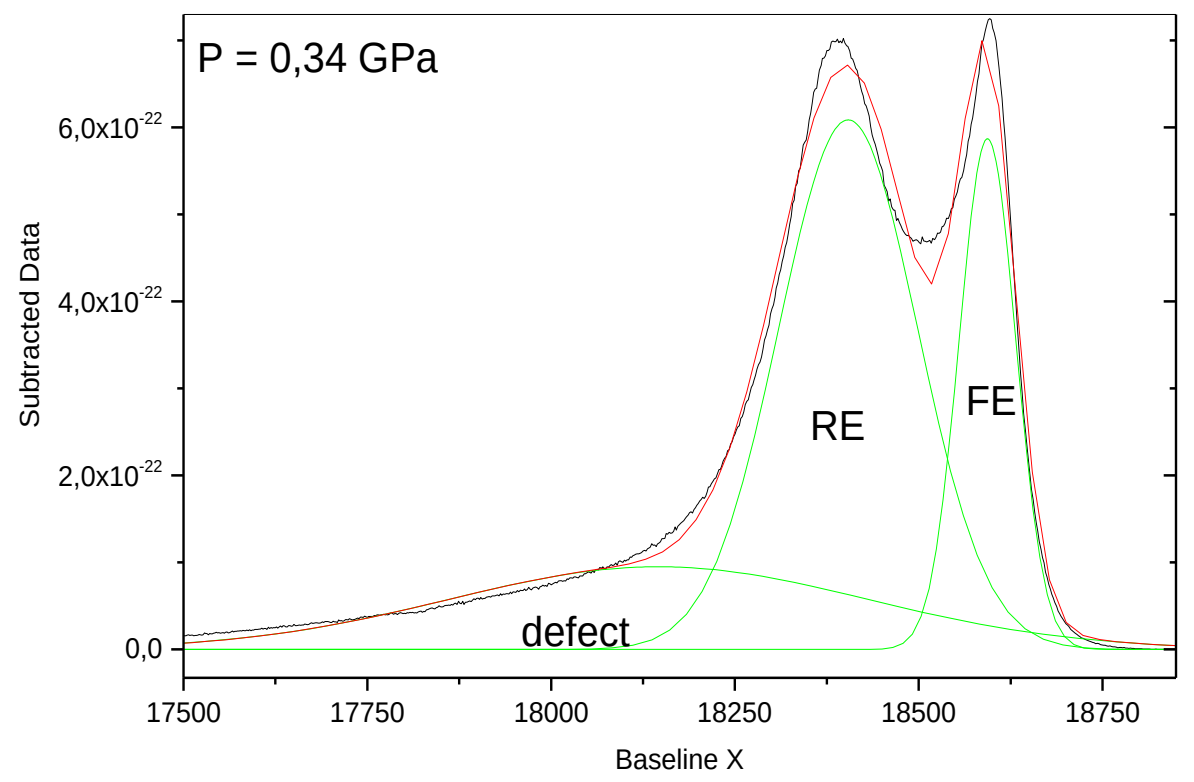

Fitting Results

| Peak Index | Peak Type | Area Intg | FWHM | Max Height | Center Grvty | Area IntgP |
|---|---|---|---|---|---|---|
| 1 | Gaussian | 5,67294E-20 | 90,79117 | 5,86992E-22 | 18593,59185 | 21,12007 |
| 2 | Gaussian | 1,44088E-19 | 222,36781 | 6,08729E-22 | 18404,08207 | 53,64329 |
| 3 | Gaussian | 6,77866E-20 | 669,80559 | 9,50743E-23 | 18146,0439 | 25,23663 |

## Peak Analysis

Data Set:[Default21]Data!C Date:10.01.2023
BaseLine:Line
Chi^2=6,20177E-047 Adj. R-Square=9,98218E-001 # of Data Points=1026
SS=6,30720E-044 Degree of Freedom=1017

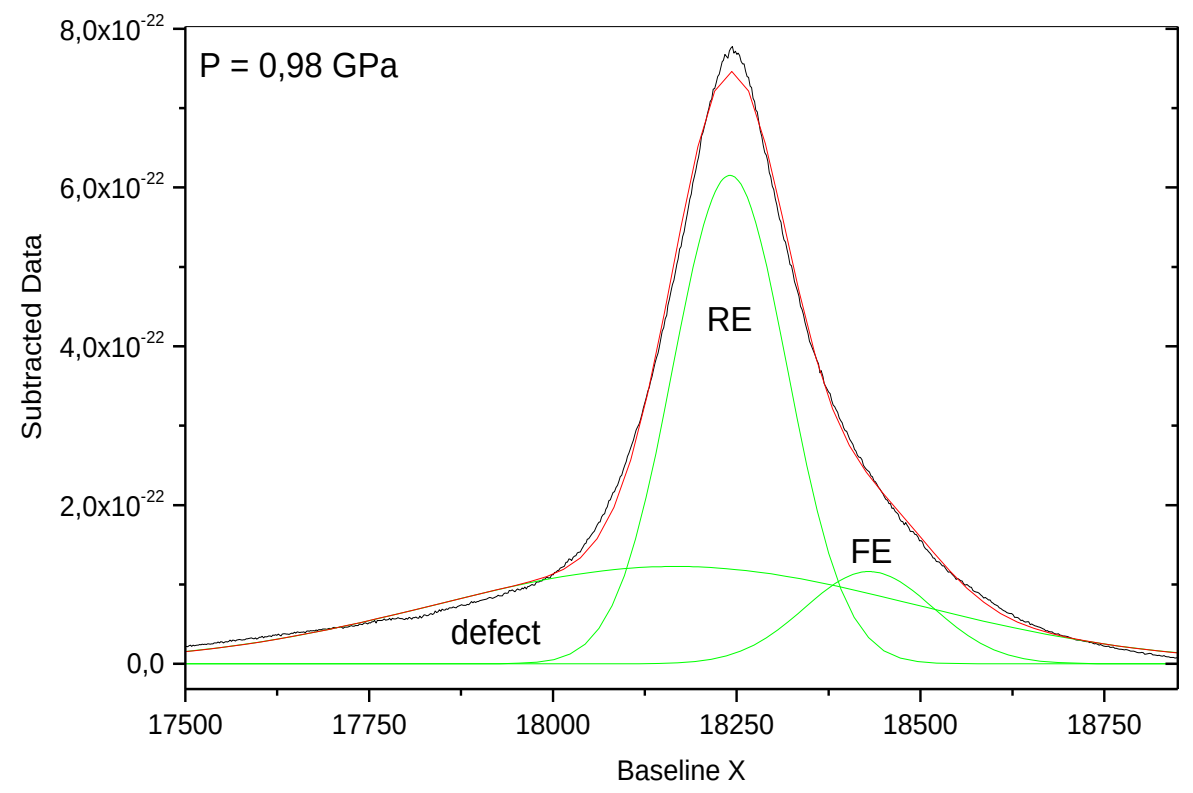

Fitting Results

| Peak Index | Peak Type | Area Intg | FWHM | Max Height | Center Grvty | Area IntgP |
|---|---|---|---|---|---|---|
| 1 | Gaussian | 2,55053E-20 | 206,2646 | 1,16165E-22 | 18429,93388 | 10,37534 |
| 2 | Gaussian | 1,19905E-19 | 183,09469 | 6,15219E-22 | 18240,9677 | 48,77631 |
| 3 | Gaussian | 1,00416E-19 | 769,36135 | 1,22614E-22 | 18166,60762 | 40,84835 |

## Peak Analysis

Data Set:[Default25]Data!C Date:10.01.2023
BaseLine:Line
Chi^2=2,89096E-047 Adj. R-Square=9,98537E-001 # of Data Points=1026
SS=2,94011E-044 Degree of Freedom=1017

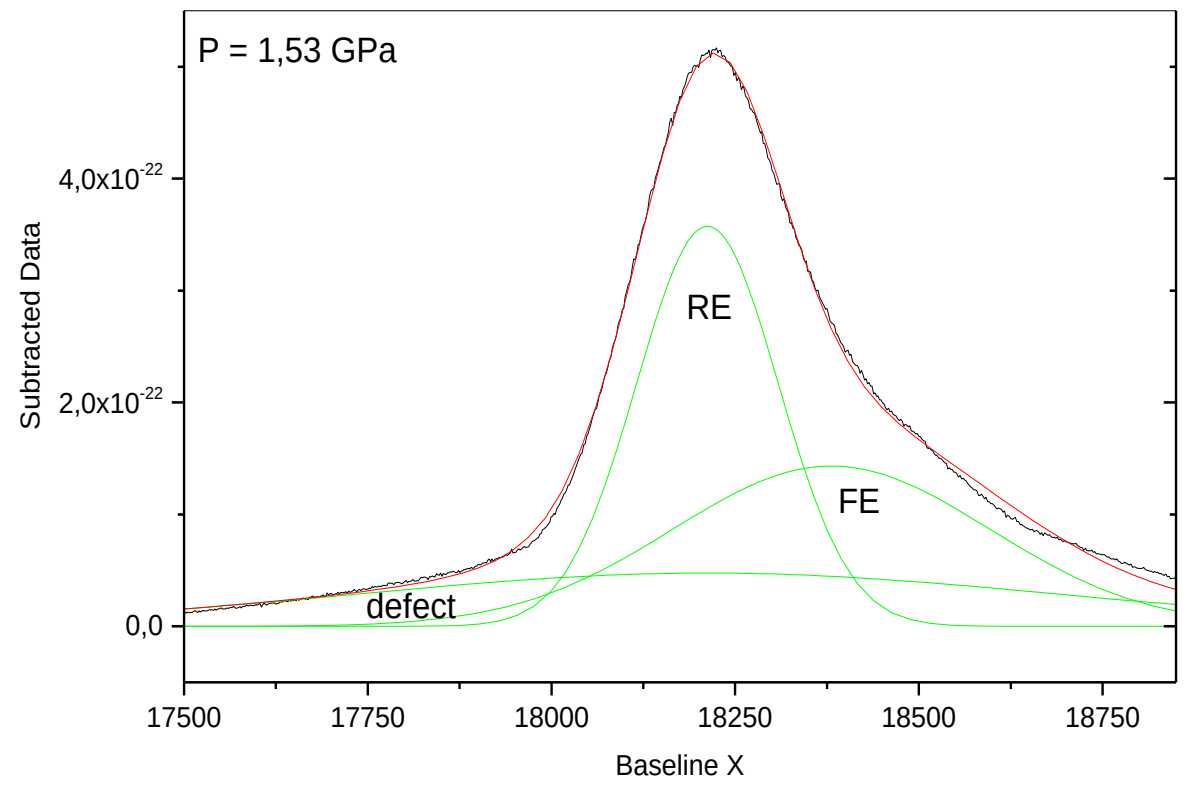

Fitting Results

| Peak Index | Peak Type | Area Intg | FWHM | Max Height | Center Grvty | Area IntgP |
|---|---|---|---|---|---|---|
| 1 | Gaussian | 7,73457E-20 | 507,33604 | 1,43221E-22 | 18380,91827 | 35,05936 |
| 2 | Gaussian | 8,64562E-20 | 227,30026 | 3,57325E-22 | 18212,29413 | 39,18898 |
| 3 | Gaussian | 5,68116E-20 | 1121,30279 | 4,75973E-23 | 18212,29413 | 25,75166 |

**Fig.SI. 6.** Fitting of some $CsPbBr_3$ high-pressure spectra (0.2, 0.34, 0.98, and 1.53 GPa).

## 5. Intensity and position of the luminescence lines as a function of pressure

If we consider all the data from both measurements (refer to Fig.SI. 4), the pressure dependence of the intensity would resemble that shown in Fig.SI. 7a. Consequently, the integral intensity was calculated by integrating under the luminescence spectrum. Subsequently, the intensity of the Rashba exciton was estimated by identifying the maximum intensity of the Rashba line. Similarly, the intensity of the free exciton was estimated as the difference between the integral intensity and the Rashba exciton intensity. This approach tends to overestimate the intensity of the free exciton, particularly in the low-pressure range (below 0.3 GPa), due to the presence of other lines in the spectra at that pressure range. Nevertheless, this approach yields qualitatively similar results to those obtained from the fitting in Fig.SI. 7 a. The fitting process involved additional data processing steps such as background subtraction, selecting the number of Gaussian functions for the fit, and occasionally manually adjusting or fixing certain fit parameters. In contrast, the second approach (Fig.SI. 7b) required less manual intervention. Considering the reduced need for manual adjustments, lower data scattering, and overall similarity of the results, the outcome of the second approach was presented in the main text of the study.

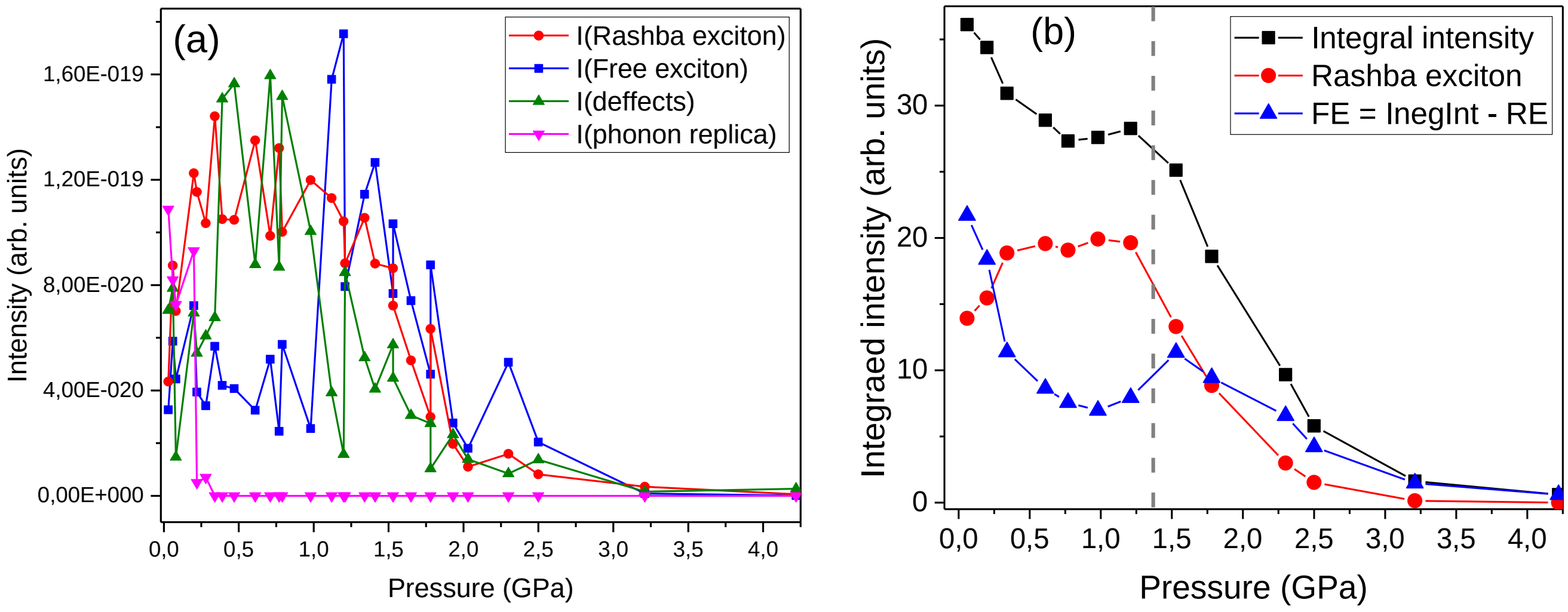

**Fig.SI. 7.** Intensity of the luminescence lines (a) – from fitting, (b) – a copy of Fig. 3 from the main text.